\documentclass[12pt]{iopart}
\usepackage{iopams}

\newtheorem{theorem}{Theorem}[section]
\newtheorem{proposition}[theorem]{Proposition}
\newtheorem{lemma}[theorem]{Lemma}

\newtheorem{definition}[theorem]{Definition}

\newtheorem{remark}[theorem]{Remark}
\newlength{\vshift}
\newlength{\hshift}
\usepackage{amssymb,amsopn}

\begin{document}
\title{$(p,q)-$deformed Fibonacci and  Lucas polynomials: characterization and Fourier integral transforms }
\author{Mahouton Norbert Hounkonnou and Sama Arjika}
\address{International Chair of Mathematical Physics
and Applications (ICMPA-UNESCO Chair), University of
Abomey-Calavi, 072 B. P.: 50 Cotonou, Republic of Benin}
\eads{\mailto{norbert.hounkonnou@cipma.uac.bj},
\mailto{rjksama2008@gmail.com}
}

\begin{abstract}
A full characterization of $(p,q)$-deformed Fibonacci  and 
Lucas polynomials is given.  These polynomials obey non-conventional three-term
 recursion relations. Their generating functions and Fourier integral transforms  are explicitly computed and discussed. Relevant  results known in the literature are examined as particular cases.
\end{abstract}

 \today
\section{Introduction}
The classical orthogonal polynomials  (COPs) and the quantum orthogonal polynomials (QOPs), also called $q-$orthogonal polynomials, constitute an interesting set of 
special functions with potential applications in physics, in probability and statistics, in approximation theory and numerical analysis, to cite a few domains where they are involved. Since the birth of quantum mechanics, the COPs also made their appearance in the bound-state wavefunctions of exactly solvable potentials.

 Depending on the set of parameters, each family  of  orthogonal polynomials occupies different levels
within the Askey hierarchy \cite{KSL}.
For instance, the classical  Hermite polynomials
$H_n(x)$ are the ground level, the Laguerre $L_n^{(\alpha)}(x)$ and Charlier $C_n(x;a)$ polynomials
 are one level higher, the Jacobi $P_n^{(\alpha,\beta)}(x)$, the Meixner $M_n(x;\beta,c)$, the Krawtchouk $K_n(x;p,N)$ and the Meixner/Pollaczek $P_n^{(\lambda)}(x;\phi)$ polynomials are two levels higher, the Hahn $Q_n(x;\alpha,\beta,N)$, the dual Hahn $R_n(\lambda(x);\gamma,\delta,N)$ polynomials, etc. are three levels higher, and so on. Besides, all orthogonal
polynomial families in this  Askey scheme are characterized by a 
set of properties:
\begin{itemize} 
\item they are solutions of second order differential or difference equations, 
\item they are generated by three-term recurrence relations,
\item they are orthogonal with respect to  weight functions, 
\item they obey  the Rodrigues-type formulas.
\end{itemize}
  The other
polynomial families which do not obey  
the above characteristic properties,   
 do not belong to the Askey $q$-scheme. 

In this work, we deal with the study of  $(p,q)-$Fibonacci 
and $(p,q)-$Lucas polynomials characterized by
non-conventional recurrence relations. Their $q-$analogs were recently introduced  by Atakishiyev et {\it al} \cite{natig}. 

The paper  is organized as follows. In Section 2, we give a  formulation of the $(p,q)-$deformed Fibonacci and  $(p,q)-$deformed Lucas polynomials. Their generating 
functions  are computed and discussed. We perform the computation of 
the associated 
Fourier integral transforms in Section 3 and end with some concluding remarks in the Section 4.
\section{$(p,q)-$deformed Fibonacci and   Lucas polynomials}
In this section, we study in details the $(p,q)-$deformed Fibonacci and
 Lucas polynomials. The corresponding generating functions are computed and discussed.
\subsection{$(p,q)-$deformed  Fibonacci polynomials} 
Start with the following definition of
$(p,q)-$analogs of Fibonacci polynomials \cite{natig,Cigl-I,Cigl-II},
\begin{eqnarray}
\label{sama:classical}
\fl \;\; F_{n+1}(x,s)=\sum_{k=0}^{\lfloor\,n/2\,\rfloor}{n-k
\atopwithdelims()k}s^{\,k}\,x^{\,n-\,2k} 
=x^{\,n}\,{}_2F_1\left(\begin{array}{c}-\frac{n}{2}\,,\frac{1-n}{2}\\-n\end{array}\Bigg|-\frac{4s}{x^2}\right),\quad n\geq0, \end{eqnarray}
given by:
\begin{definition}
\begin{eqnarray}
\label{pippo}
F_{n+1}(x,s|p,q):= \sum_{k=0}^{\lfloor\,n/2\,\rfloor\,}(pq)^{k(k+1)
/2}\,{n-k\atopwithdelims[]k }_{p,q} s^{k}x^{n-2k},
\end{eqnarray}
where the $(p,q)-$binomial
coefficients ${\,n\,\atopwithdelims []\,k\,}_{p,q}$  are given by  \cite{Janga}
\begin{eqnarray}
\fl \quad\quad {\,n\,\atopwithdelims []\,k\,}_{p,q} :=\frac{((p,q);(p,q))_n}
{((p,q);(p,q))_k((p,q);(p,q))_{n-k}},
\end{eqnarray}
and  $(p,q)-$shifted factorial  $((a,b);(p,q))_n$ is  defined as 
\begin{eqnarray}
((a,b);(p,q))_n:=\prod_{k=0}^{n-1}(a p^k-b q^k)  \; \mbox{ for }\; n\geq 1,\quad ((a,b);(p,q))_0:=1.
\end{eqnarray}
\end{definition}
When $n\to \infty,$
\begin{eqnarray}
\big((a,b);(p,q)\big)_\infty:=\prod_{k=0}^{\infty}(a p^k-b q^k).
\end{eqnarray}
\begin{remark}
As expected, when the parameter $p\to 1$, the $(p,q)-$Fibonacci polynomials (\ref{pippo}) are reduced to their $q-$version \cite{natig}, i.e
\begin{eqnarray}
 \label{pippo2}
  F_{n+1}(x,s|q)= \sum_{k=0}^{\lfloor\,n/2\rfloor\,}q^{k(k+1)
/2}{n-k\atopwithdelims[]k }_q s^{k}x^{n-2k},
 \end{eqnarray}
where the $q-$binomial
coefficient ${\,n\,\atopwithdelims []\,k\,}_q$ is given by
\begin{eqnarray}
{\,n\,\atopwithdelims []\,k\,}_q :=\frac{(q;q)_n}
{(q;q)_k(q;q)_{n-k}},        \label{pippo3a}               
\end{eqnarray}
and $(z;q)_n$ is the $q$-shifted factorial defined as
\begin{eqnarray}
\fl \qquad (z;q)_0:=1,\quad 
(z;q)_n:=\prod_{k=0}^{n-1}(1-z q^k),\; n\geq 1,\;
(z;q)_\infty:=\prod_{k=0}^{\infty}(1-z q^k).
\end{eqnarray}
\end{remark}
%
Further, the following statement holds.
\begin{proposition}
\label{ppp}
The $(p,q)-$Fibonacci polynomials (\ref{pippo}) can be rewritten in terms of the hypergeometric function ${}_{8}\varphi_5$ as follows:
\begin{eqnarray}
\label{olll}
F_{n+1}(x,s|p,q)&=&
x^{n}{}_{8}\varphi_5\left(\begin{array}{c}(p^{-\frac{n}{2}},q^{-\frac{n}{2}}),(p^{\frac{1-n}{2}},q^{\frac{1-n}{2}}),(p^{-\frac{n}{2}},-q^{-\frac{n}{2}}),
 \\\\\nonumber
(p^{-n},q^{-n}),(p,0),(p,0),
 \end{array}\right.
\\\\\cr
&&\left.\begin{array}{c}(p^{\frac{1-n}{2}},-q^{\frac{1-n}{2}}),0,0,0,0
 \\\\\nonumber
(p,0),(p,0) \end{array}\Bigg|(p,q);-\frac{sq^{n}p^{4+n}}
{x^2}\right),    \end{eqnarray}
where the $(p,q)-$hypergeometric function ${}_{r}\varphi_s$   is defined as \cite{Janga}
\begin{eqnarray}\label{pqhyper}
{}_{r}\varphi_s\left(\begin{array}{c}(a_1,b_1),(a_2,b_2),\ldots, (a_r,b_r)
 \\
(c_1,d_1),(c_2,d_2),\ldots, (c_s,d_s)
 \end{array}\Bigg|
(p,q);x\right) \cr
 :=\sum_{n=0}^\infty\frac{((a_1,b_1)(a_2,b_2)\ldots(a_r,b_r);(p,q))_n}{((c_1,d_1)(c_2,d_2)\ldots(c_s,d_s);(p,q))_n}\frac{\Big[(-1)^n(q/p)^{({}^n_2)}\Big]^{1+s-r}}{((p,q);(p,q))_n} x^n.
\end{eqnarray}
\end{proposition}
{\bf Proof.} By using the  $(p,q)-$identities:
\begin{eqnarray*}
\fl \qquad ((p,q);(p,q))_{n-k}=&\frac{((p,q);(p,q))_n}{((p^{-n},q^{-n});(p,q))_k}(-1)^k(pq)^{({}^k_2)-nk},\\
\fl \qquad ((p^n,q^n);(p,q))_{2k}
=&\big((p^{\frac{n}{2}},q^{\frac{n}{2}}),(p^{\frac{1+n}{2}},q^{\frac{1+n}{2}}),(p^{\frac{n}{2}},-q^{\frac{n}{2}}),(
p^{\frac{1+n}{2}},-q^{\frac{1+n}{2}});(p,q)\big)_k,
\end{eqnarray*}
we  get
\begin{eqnarray*}
F_{n+1}(x,s|p,q)
&=&\sum_{k=0}^{\lfloor\,n/2\,\rfloor\,}\frac{(pq)^{k(k+1)
/2}((p,q);(p,q))_{n-k}}
{((p,q);(p,q))_k((p,q);(p,q))_{n-2k}}s^{k}x^{n-2k}\cr
&=&x^n\sum_{k=0}^{\lfloor\,n/2\,\rfloor\,}\frac{(p^{-1}q)^{-2({}^k_2)}((p^{-n},q^{-n});(p,q))_{2k}}
{((p^{-n},q^{-n}),(p,q);(p,q))_kp^{4({}^{k+1}_{\;\;2})}}\Bigg(-\frac{sq^np^{n+4}}{x^2}\Bigg)^k,
\end{eqnarray*}
with $p^{({}^{k+1}_{\;\;2})}=((p,0);(p,q))_k$. 
$\square$
\begin{remark}
In the limit when $p\to 1$, (\ref{pqhyper}) is reduced to $q-$hypergeometric function  characterizing 
the $q-$Fibonacci polynomials investigated in \cite{natig}, i. e.
\begin{eqnarray}
\fl \qquad F_{n+1}(x,s|q)=x^{n}{}_4\phi_1\left(\begin{array}{c}
q^{-n/2},q^{(1-n)/2},-q^{-n/2}
,-q^{(1-n)/2}\\
q^{-n}\end{array}\Bigg|q;-\frac{q^{n}s}
{x^2}\right), n\geq0.
 \end{eqnarray}
\end{remark}
\begin{lemma}
\label{sama:lem}
The $(p,q)-$binomials coefficients 
\begin{eqnarray}
{n-k\atopwithdelims[]k }_{p,q}=\frac{((p,q);(p,q))_{n-k}}
{((p,q);(p,q))_k((p,q);(p,q))_{n-2k}},
\end{eqnarray}
where $0\leq 2k\leq n,\;n\in\mathbb{N}$ satisfy the following identities:
\begin{eqnarray}
\label{sum}
{\,n-k\,\atopwithdelims []\,k\,}_{p,q}=q^k{\,n-1-k\,\atopwithdelims []\,k\,}_{p,q}+p^{n-2k}{\,n-1-k\,\atopwithdelims []\,k-1\,}_{p,q},
\end{eqnarray}
\begin{eqnarray}
\label{sume}
{\,n-k\,\atopwithdelims []\,k\,}_{p,q}=p^k{\,n-1-k\,\atopwithdelims []\,k\,}_{p,q}+q^{n-2k}{\,n-1-k\,\atopwithdelims []\,k-1\,}_{p,q},
\end{eqnarray}
\begin{eqnarray}
\label{sumer}
 {\,n-k\,\atopwithdelims []\,k\,}_{p,q}&=&p^k{\,n-1-k\,\atopwithdelims []\,k\,}_{p,q}+p^{n-k}q^{-k}{\,n-1-k\,\atopwithdelims []\,k-1\,}_{p,q}\cr
&-&(p^{n-2k+1}-q^{n-2k+1})q^{-k}{\,n-k\,\atopwithdelims []\,k-1\,}_{p,q}.
\end{eqnarray}
\end{lemma}
{\bf Proof.}
Using the relations \cite{Cigl-I}:
\begin{eqnarray}
\fl \qquad  {\,n-k\,\atopwithdelims []\,k\,}_{q}=q^k{\,n-1-k\,\atopwithdelims []\,k\,}_{q}+{\,n-1-k\,\atopwithdelims []\,k-1\,}_{q},\quad {\,n\,\atopwithdelims []\,k\,}_{q/p}=p^{-k(n-k)}{\,n\,\atopwithdelims []\,k\,}_{p,q},\end{eqnarray}
\begin{eqnarray}
\fl \qquad 
{\,n-k\,\atopwithdelims []\,k\,}_{q}={\,n-1-k\,\atopwithdelims []\,k\,}_{q}+q^{n-2k}{\,n-1-k\,\atopwithdelims []\,k-1\,}_{q}, \end{eqnarray}
and 
\begin{eqnarray}
\fl \qquad q^k{\,n-k\,\atopwithdelims []\,k\,}_{q}=q^k{\,n-1-k\,\atopwithdelims []\,k\,}_{q}+{\,n-1-k\,\atopwithdelims []\,k-1\,}_{q}-(1-q^{n-2k+1}){\,n-k\,\atopwithdelims []\,k-1\,}_{q},
\end{eqnarray}
yields the required identities:
\begin{eqnarray}
{\,n-k\,\atopwithdelims []\,k\,}_{p,q}=q^k{\,n-1-k\,\atopwithdelims []\,k\,}_{p,q}+p^{n-2k}{\,n-1-k\,\atopwithdelims []\,k-1\,}_{p,q},
\end{eqnarray}
\begin{eqnarray}
{\,n-k\,\atopwithdelims []\,k\,}_{p,q}=p^k{\,n-1-k\,\atopwithdelims []\,k\,}_{p,q}+q^{n-2k}{\,n-1-k\,\atopwithdelims []\,k-1\,}_{p,q},
\end{eqnarray}
and 
\begin{eqnarray}
 {\,n-k\,\atopwithdelims []\,k\,}_{p,q}&=&p^k{\,n-1-k\,\atopwithdelims []\,k\,}_{p,q}+p^{n-k}q^{-k}{\,n-1-k\,\atopwithdelims []\,k-1\,}_{p,q}\cr
&-&(p^{n-2k+1}-q^{n-2k+1})q^{-k}{\,n-k\,\atopwithdelims []\,k-1\,}_{p,q}.
\end{eqnarray}
$\square$
\begin{proposition}
\label{por}
The $(p,q)-$deformed Fibonacci polynomials satisfy the   
following non-standard three-term recursion relations:
\begin{eqnarray}
\label{a1re}
 \fl \qquad F_{n+1}(x,s|p,q) &=& x F_{n}
(x,q s|p,q)+ s q p^{n-1}F_{n-1}(x,q p^{-1}s|p,q),\\
\label{aee2}
\fl  &=& x F_{n}
(x,s p|p,q)+ s p q^{n-1}F_{n-1}(x,s p q^{-1}|p,q),  \\   
\label{aae2}
 \fl  &=& \big(x+ s p(q-p)D_{(p,q)}\big)F_{n}
(x,s p|p,q)
+s p^{n}F_{n-1}(x,s|p,q),   \, n\geq 1,
   \end{eqnarray} 
with the initial conditions $F_{0}(x,s|p,q)=0$, $F_{1}(x,s|p,q)=1$ and 
  the $(p,q)-$Jackson's derivative $D_{(p,q)}$   given by \cite{Janga}
\begin{eqnarray}
D_{(p,q)}f(x)=\frac{f(p x)-f(q x)}{(p-q)x}.
 \end{eqnarray}
\end{proposition}
{\bf Proof.} By multiplying the equations (\ref{sum}), (\ref{sume}) and (\ref{sumer}) of the 
Lemma \ref{sama:lem} by $(p q)^{k(k+1)/2}s^{k}x^{n-2k}$ and
summing from $k=0$ to $\lfloor\,n/2\,\rfloor$, we get 
\begin{eqnarray*}
 \fl \qquad \sum_{k=0}^{\lfloor\,n/2\,\rfloor\,}(p q)^{k(k+1)/2}{\,n-k\,\atopwithdelims []\,k\,}_{p,q}s^{k}x^{n-2k}= \sum_{k=0}^{\lfloor\,n/2\,\rfloor\,}(p q)^{k(k+1)/2}{\,n-1-k\,\atopwithdelims []\,k\,}_{p,q}(q s)^{k}x^{n-2k}\cr
+ p^{n}\sum_{k=1}^{\lfloor\,n/2\,\rfloor\,}(p q)^{k(k+1)/2}{\,n-1-k\,\atopwithdelims []\,k-1\,}_{p,q}(p^{-2}s)^{k}x^{n-2k},
\end{eqnarray*}
\begin{eqnarray*}
 \fl \qquad \sum_{k=0}^{\lfloor\,n/2\,\rfloor\,}(p q)^{k(k+1)/2}{\,n-k\,\atopwithdelims []\,k\,}_{p,q}s^{k}x^{n-2k}= \sum_{k=0}^{\lfloor\,n/2\,\rfloor\,}(p q)^{k(k+1)/2}{\,n-1-k\,\atopwithdelims []\,k\,}_{p,q}(p s)^{k}x^{n-2k}\cr
+q^{n}\sum_{k=1}^{\lfloor\,n/2\,\rfloor\,}(p q)^{k(k+1)/2}{\,n-1-k\,\atopwithdelims []\,k-1\,}_{p,q}(q^{-2}s)^{k}x^{n-2k},
\end{eqnarray*}
and 
\begin{eqnarray*}
 \fl \qquad \sum_{k=0}^{\lfloor\,n/2\,\rfloor\,}(p q)^{k(k+1)/2}{\,n-k\,\atopwithdelims []\,k\,}_{p,q}s^{k}x^{n-2k}= \sum_{k=0}^{\lfloor\,n/2\,\rfloor\,}(p q)^{k(k+1)/2}{\,n-1-k\,\atopwithdelims []\,k\,}_{p,q}(p s)^{k}x^{n-2k}\cr
+ p^{n}\sum_{k=1}^{\lfloor\,n/2\,\rfloor\,}(p q)^{k(k+1)/2}{\,n-1-k\,\atopwithdelims []\,k-1\,}_{p,q}(p^{-1}q^{-1}s)^{k}x^{n-2k}\cr
+ p^{n}\sum_{k=1}^{\lfloor\,n/2\,\rfloor\,}(p q)^{k(k+1)/2}{\,n-k\,\atopwithdelims []\,k-1\,}_{p,q}(p^{n-2k+1}-q^{n-2k+1})(q^{-1}s)^{k}x^{n-2k},
\end{eqnarray*}
which are  equivalent to
\begin{eqnarray}
F_{n+1}(x,s|p,q) = x F_{n}
(x,q s|p,q)+ s q p^{n-1}F_{n-1}(x,q p^{-1}s|p,q),
\\
F_{n+1}(x,s|p,q) = x F_{n}
(x,s p|p,q)+ s p q^{n-1}F_{n-1}(x,s p q^{-1}|p,q),   
\end{eqnarray}
and 
\begin{eqnarray}
\fl \qquad  F_{n+1}(x,s|p,q) = \big(x+ s p(q-p)D_{(p,q)}\big)F_{n}
(x,s p|p,q)
+s p^{n}F_{n-1}(x,s|p,q),   \, n\geq 1,
\end{eqnarray}
respectively, with $F_{0}(x,s|p,q)=0$, $F_{1}(x,s|p,q)=1$.
 $\square$
\begin{remark}
In the limit case when $p\to 1$, the equations (\ref{a1re})-(\ref{aae2}) are reduced to their $q-$analogs \cite{natig,Cigl-I}, i.e
\begin{eqnarray}
\label{a1}
\fl \qquad F_{n+1}(x,s|q) &=& x F_{n}
(x,q s|q)+ s q F_{n-1}(x,q s|q),\\
\label{a2}
 \fl &=& x F_{n}
(x,s|q)+ sq^{n-1}F_{n-1}(x,sq^{-1}|q),\\
\label{ae2}
  \fl &=& x F_{n}
(x,s|q)+ s(q-1)D_qF_{n}
(x,s|q)+s F_{n-1}(x,s|q),
\; n\geq1,      
\end{eqnarray}
with the initial values $F_{0}(x,s|q)=0$ and 
$F_{1}(x,s|q)=1,$ where $D_q$
is the $q-$Jackson differential 
operator defined by
\begin{eqnarray}
 D_qf(x):=\frac{f(x)-f(qx)}{(1-q)x}.
\end{eqnarray}
\end{remark}
\begin{definition}
The  
generating function $f_F(x,s;t|p,q)$ associated
with the $(p,q)$-Fibonacci polynomials is defined as follows:
\begin{eqnarray}
 \label{pippo4} 
f_F(x,s;t|p,q):=\sum_{n=0}^{\infty}F_{n}(x,s p^{-n}|p,q)t^{n}.
\end{eqnarray}
\end{definition}
\begin{proposition}
\label{ropo}
The generating functions 
 (\ref{pippo4}) can be re-expressed in terms of the hypergeometric function ${}_2\varphi_2:$
\begin{eqnarray}
 \label{pip} 
f_F(x,s;t|p,q)= \frac{t}{1-x t}\,{}_2\varphi_2\left(\begin{array}{l}(p,q),\quad0\\
(p,x t q),(p,0)\end{array}
\Bigg|(p,q);-q st^2\right), \;|t|<1.
\end{eqnarray}
\end{proposition}
{\bf Proof.} 
From (\ref{pippo}), we have
\begin{eqnarray}
\label{sama:poro}
\fl  f_F(x,s;t|p,q):&=&\sum_{n=0}^{\infty}F_{n}(x,s p^{-n}|p,q)\,t^{n}\cr
\fl &=&\sum_{n=0}^{\infty} \sum_{k=0}^{\lfloor\,\frac{n-1}{2}\,\rfloor\,}
(p q)^{k(k+1)/2}{\,n-k\,\atopwithdelims []\,k\,}_{p,q}s^{k}p^{-n k}x^{n-2k}t^{n}\cr
\fl &=&\sum_{k=0}^{\infty}\frac{(p q)^{k(k+1)/2}s^k}{((p,q);(p,q))_k} 
 \sum_{n-1=2k}^{\infty}\frac{((p,q);(p,q))_{n-1-k}}{((p,q);(p,q))_{n-1-2k}}x^{n-1-2k}t^n p^{-n k}
\cr
\fl &=&\sum_{k=0}^{\infty}\frac{(p q)^{k(k+1)/2}s^k}{((p,q);(p,q))_k} 
 \sum_{m=0}^{\infty}\frac{((p,q);(p,q))_{m+k}}{((p,q);(p,q))_{m}}x^{m}t^{m+1+2k}p^{-(m+1+2k)k}
\cr
\fl &=&t\sum_{k=0}^{\infty}\frac{(p q)^{k(k+1)/2}p^{-2k^2-k}(st^2)^k}{((p,q);(p,q))_k} \cr
 &\times&\sum_{m=0}^{\infty}\frac{((p,q);(p,q))_k((p^{1+k},q^{1+k});(p,q))_m}{((p,q);(p,q))_{m}}(x t p^{-k})^m
\cr
\fl &=&t\sum_{k=0}^{\infty} (p q)^{k(k+1)/2}p^{-2k^2-k}(st^2)^k 
 \sum_{m=0}^{\infty}\frac{ ((p^{1+k},q^{1+k});(p,q))_m}{((p,q);(p,q))_{m}}(x t p^{-k})^m
\cr
\fl &=&t\sum_{k=0}^{\infty} (p q)^{k(k+1)/2}p^{-2k^2-k}(st^2)^k 
{}_1\varphi_0\left(\begin{array}{c}(p^{1+k},q^{1+k})\\
-\end{array}
\Bigg|(p,q); x t p^{-k}\right)
\cr
\fl &=&t\sum_{k=0}^{\infty} (p q)^{k(k+1)/2}p^{-2k^2-k}(st^2)^k 
\frac{((p,x t p^{-k}q^{1+k});(p,q))_\infty}{((p,x t p);(p,q))_\infty}.
\end{eqnarray}
By using the equation (50) of \cite{Janga}, the expression (\ref{sama:poro}) is transformed into 
\begin{eqnarray*}
f_F(x,s;t|p,q)&=&t\sum_{k=0}^{\infty} (p q)^{k(k+1)/2}p^{-2k^2-k}(st^2)^k 
\frac{p^{({}^{k+2}_{\;\;2})}}{((p,x t p);(p,q))_{k+1}}
\cr
&=&\frac{t}{1-xt}\sum_{k=0}^{\infty}
\frac{ (p^{-1}q)^{k(k-1)/2}(st^2q)^k }{((p,x t q),(p,0);(p,q))_k},
\end{eqnarray*}
which achieves the proof.  $\square$
\begin{remark}
In the limit when $p\to1$, the generating function (\ref{pip}) is reduced to its $q-$version provided  
by Atakishiyev et {\it al}  \cite{natig}:
\begin{eqnarray}
\label{geneq}
\fl \quad \qquad  f_F(x,s;t|q):= \sum_{n=0}^{\infty}F_{n}(x,s |q)t^{n}
= \frac{t}{1-x t}\,{}_1\phi_1\left(\begin{array}{c}q\\
q x t\end{array}
\Bigg|q;- q st^2 \right),\; |t|<1.  
 \end{eqnarray}
\end{remark}
\begin{remark}
In the limit
\begin{enumerate}
\item When $p,q\to 1$, the $(p,q)-$deformed Fibonacci polynomials (\ref{pippo}) are reduced to
the classical case $F_n(x,s)$
 given by the explicit sum formula  \cite{Cigl-I,Cigl-II}
\begin{eqnarray}
\label{sama:classical}
\fl \;\; F_{n+1}(x,s)=\sum_{k=0}^{\lfloor\,n/2\,\rfloor}{n-k
\atopwithdelims()k}s^{\,k}\,x^{\,n-\,2k} 
=x^{\,n}\,{}_2F_1\left(\begin{array}{c}-\frac{n}{2}\,,\frac{1-n}{2}\\-n\end{array}\Bigg|-\frac{4s}{x^2}\right),\quad n\geq0, \end{eqnarray}
where ${\,n\,\atopwithdelims()\,k\,}:= n!/[k!(n-k)!]$ is the binomial
coefficient and ${}_2F_1$ is a hypergeometric function \cite{ASK}.
They obey  the following three-term recursion
relation
\begin{eqnarray}
\label{ttrsama}
F_{n+1}(x,s) = x F_{n}(x,s)+s F_{n-1}(x,s),\quad
n\geq1\,,                                                  \end{eqnarray}
with initial values $F_0(x,s)=0$ and $F_1(x,s)=1$.
Their generating function $f_F(x,s;t)$ is given by \cite{ASK}
\begin{eqnarray}
\label{sama:teg}
f_F(x,s;t):=\,\sum_{n=0}^{\infty}\,F_{n}(x,s)\,t^{\,n}\,=\,\frac
{t}{1-x\,t-s\,t^{\,2}},\quad |\,t\,|<1.           \end{eqnarray}
The polynomials (\ref{sama:classical}) are monic and normalized so that
for $s=1$, one recovers the Fibonacci polynomials $f_n(x)=F_n(x,1)$
introduced by Catalan  (\cite{Koshy}, eq. (37.1) p. 443). 
\item When $x=s=1$, the $(p,q)-$Fibonacci polynomials (\ref{pippo}) 
furnish the $(p,q)-$Fibonacci number $F_n(1,1|p,q):=F_n(p,q)$ given by 
\begin{eqnarray}
F_n(p,q)=\sum_{k=0}^{\lfloor\,n/2\,\rfloor\,}(pq)^{k(k+1)
/2}\,{n-k\atopwithdelims[]k }_{p,q},
\end{eqnarray}
which is the $(p,q)-$extension of the $q-$Fibonacci number \cite{Cigl-II} 
\begin{eqnarray}
F_n(q)=\sum_{k=0}^{\lfloor\,n/2\,\rfloor\,}q^{k(k+1)
/2}\,{n-k\atopwithdelims[]k }_{q}.
\end{eqnarray}
From the limit $(p,q)\to (1, 1)$, we recover the  classical Fibonacci number, i.e.
\begin{eqnarray}
\label{sama:numbers}
  F_n=\frac{1}{2^{n-1}}\sum_{k=0}^{\lfloor\,\frac{n-1}{2}\,\rfloor}{n
\atopwithdelims()2k+1}5^k.
\end{eqnarray} 
The $(p,q)-$deformed generating function associated  with the $(p,q)-$Fibonacci number is given by
\begin{eqnarray}
f_F(t|p,q)= \frac{t}{1- t}\,{}_2\varphi_2\left(\begin{array}{l}(p,q),\quad0\\
(p, t q),(p,0)\end{array}
\Bigg|(p,q);-q t^2\right), \;|t|<1
\end{eqnarray}
which is reducible to the following corresponding $q-$generating function by passing to the limit  $p\to1:$
\begin{eqnarray}
f_F(t|q)= \frac{t}{1- t}\,{}_1\phi_1\left(\begin{array}{c}q\\
q t\end{array}
\Bigg|q;- q t^2 \right), \;|t|<1. 
\end{eqnarray}
 Finally, the limit case when $(p,q)\to (1, 1)$ yields 
the classical 
version of the generating function $f_F(t)$
of the Fibonacci polynomials (see \cite{Dunlap, NalliHauk, StakhAran,Bers} for more details), i.e.
\begin{eqnarray}
f_F(t)= \sum_{n=0}^{\infty}\,F_{n} t^{\,n}\,=\,\frac
{t}{1- t- t^{ 2}},\quad |\,t\,|<1.          
\end{eqnarray}
\end{enumerate}
\end{remark}
\subsection{ $(p,q)-$deformed Lucas polynomials}
Let us introduce the $(p,q)-$analogs of the Lucas polynomials: 
\begin{eqnarray}
\fl\quad L_{n}(x,s):= \sum_{k=0}^{\lfloor\,n/2\,\rfloor}
\frac{n}{n-k}\,{n-k\atopwithdelims()k}
 s^{k}x^{n-2k}=x^{\,n}\,{}_2F_1\left(\begin{array}{c}-\frac{n}{2}\,,\frac{1-n}{2}
\\1-n\end{array}\Bigg|-\frac{4s}{x^2}\right),\; n\geq0, 
\end{eqnarray}
as follows:
\begin{definition}
\begin{eqnarray}
 \label{pippo8}
L_{n}(x,s|p,q):= \sum_{k=0}^{\lfloor\,n/2\,\rfloor}
(p q)^{ ({}^k_2)}\frac{[n]_{p,q}}{[n-k]_{p,q}}\,{n-k\atopwithdelims[]k}
_{p,q} s^{k}x^{n-2k}, 
\end{eqnarray}
where the $(p,q)-$number $[n]_{p,q}$ is given by
\begin{eqnarray}
[n]_{p,q}:=\frac{p^n-q^n}{p-q}.
\end{eqnarray}
\end{definition}
In the limit when $p\to1$, the $(p,q)-$Lucas polynomials are reduced to 
the $q-$Lucas polynomials introduced in \cite{Cigl-I} 
\begin{eqnarray}
L_{n}(x,s|q):= \sum_{k=0}^{\lfloor\,n/2\,\rfloor}
q^{ ({}^k_2)}\frac{[n]_{q}}{[n-k]_{q}}\,{n-k\atopwithdelims[]k}
_{q} s^{k}x^{n-2k}, 
\end{eqnarray}
where the $q-$number $[n]_{q}$ is defined as
\begin{eqnarray}
[n]_{q}:=\frac{1-q^n}{1-q}.
\end{eqnarray}
The following proposition holds.
\begin{proposition}
The $(p,q)-$Lucas polynomials (\ref{pippo8}) can be defined as follows:
\begin{eqnarray}
\label{caslu}
L_n(x,s|p,q)&=&
x^{n}{}_{8}\varphi_5\left(\begin{array}{c}(p^{-\frac{n}{2}},
q^{-\frac{n}{2}}),(p^{\frac{1-n}{2}},q^{\frac{1-n}{2}}),(p^{-\frac{n}{2}},-q^{-\frac{n}{2}}),
 \\\\\nonumber
(p^{1-n},q^{1-n}),(p,0),(p,0),
 \end{array}\right.
\\\\\cr
&&\left.\begin{array}{c}(p^{\frac{1-n}{2}},-q^{\frac{1-n}{2}}),0,0,0,0
 \\\\\nonumber
(p,0),(p,0) \end{array}\Bigg|(p,q);-\frac{sq^{n}p^{4+n}}
{x^2}\right).    \end{eqnarray}
\end{proposition}
{\bf Proof.} The proof is the same as in the Proposition \ref{ppp}. $\square$
\begin{lemma}
The $(p,q)-$coefficients 
\begin{eqnarray}
\frac{[n]_{p,q}}{[n-k]_{p,q}}{\,n-k\,\atopwithdelims []\,k\,}_{p,q}
\end{eqnarray}
satisfy the following identities: 
\begin{eqnarray}
\label{asumm}
\frac{[n]_{p,q}}{[n-k]_{p,q}}{\,n-k\,\atopwithdelims []\,k\,}_{p,q}=
q^k{\,n-k\,\atopwithdelims []\,k\,}_{p,q}+p^{n-2k}{\,n-1-k\,\atopwithdelims []\,k-1\,}_{p,q} 
\end{eqnarray}
and 
\begin{eqnarray}
\frac{[n]_{p,q}}{[n-k]_{p,q}}{\,n-k\,\atopwithdelims []\,k\,}_{p,q}=
p^{k}{\,n-k\,\atopwithdelims []\,k\,}_{p,q}+q^{n-k}{\,n-1-k\,\atopwithdelims []\,k-1\,}_{p,q}. 
\end{eqnarray}
\end{lemma}
Besides, in analogous way as for the Fibonacci 
polynomials, we can prove the following result.
\begin{proposition}
The $(p,q)-$Lucas polynomials (\ref{pippo8}) satisfy   
non-standard  recursion relations for $n\geq 1:$
\begin{eqnarray}
\label{tas}
&&L_{n}(x,s|p,q) =  F_{n+1}
(x,p^{-1} s|p,q)+ s p^{n-1} F_{n-1}(x, s p^{-1}|p,q),\\
\label{seza2}
&&L_{n}(x,sqp^{-1}|p,q) =   F_{n+1}
(x,s p^{-1}|p,q)+ s p^{-1}q^{n}F_{n-1}(x,p^{-1}s|p,q)
\end{eqnarray}
with $L_{0}(x,s|p,q)=1$, $L_{1}(x,s|p,q)=x$.
\end{proposition}
\begin{remark}
In the limit when $p\to 1$,  the  polynomials (\ref{pippo8}) are reduced to the
well known $q-$Lucas polynomials
$L_{n}(x,s|q)$ studied by Atakishiyev  {\it  et al} \cite{natig}: 
\begin{eqnarray}
 \label{definitionlucas}
\fl \quad L_{n}(x,s|q):&=&\sum_{k=0}^{\lfloor n/2 \rfloor}
q^{k(k-1)/2}\frac{[n]_q}{[n-k]_q}{n-k\atopwithdelims[]k}
_q s^{k}x^{n-2k}\cr
\fl &=&x^{n}{}_4\phi_1\left(\begin{array}{c}q^{-n/2},q^{(1-n)/2},-q^{-n/2}
,-q^{(1-n)/2}\\
q^{1-n}\end{array}\Bigg|q;-\frac{q^{n}s}{x^2}
\right), \;n\geq 0, 
\end{eqnarray}
where the $q-$number $[n]_q$ is defined as 
$
[n]_q:=(1-q^n)/(1-q),
$
 satisfying the following recursion relations
\begin{eqnarray}
\label{aas}
&&L_{n}(x,s|q) =  F_{n+1}
(x,s|q)+ s F_{n-1}(x, s|q),
\\
\label{sza2}
&&L_{n}(x,sq|q) =   F_{n+1}
(x,s|q)+ sq^{n}F_{n-1}(x,s|q)
\end{eqnarray}
with $L_{0}(x,s|q)=1,\;L_{1}(x,s|q)=x.$
\end{remark}
The proof is similar to that previously performed  for the Fibonacci polynomials.
\begin{definition}
The  
generating function  $f_L(x,s;t|p,q)$ associated
with the $(p,q)-$Lucas  polynomials $L_n(x,s|p,q)$  is defined by
\begin{eqnarray}
\label{eaaa}
f_L(x,s;t|p,q):=\sum_{n=0}^{\infty}L_{n}(x,s p^{-n}|p,q)t^{n}. 
\end{eqnarray}
\end{definition}
\begin{proposition}
The generating functions 
(\ref{eaaa}) is explicitly given by
\begin{eqnarray}
\label{ase}
\fl \qquad \qquad f_L(x,s;t|p,q)=
\frac{ 1+s p t^2 }{1-x p t}\,{}_2\varphi_2\left(\begin{array}{l}(p,q),\quad0\\
(p,x t p q),(p,0)\end{array}
\Bigg|(p,q);-q st^2\right), \;|t|<1.
\end{eqnarray}
\end{proposition}
{\bf Proof.} The proof is immediate from  the definition:
\begin{eqnarray}
\fl \quad f_L(x,s;t|p,q):&=&\sum_{n=0}^{\infty}L_{n}(x,s p^{-n}|p,q)t^{n}\cr
&=&\sum_{n=0}^{\infty}F_{n+1}
(x,p^{-1-n} s|p,q)t^{n}+ s \sum_{n=0}^{\infty} F_{n-1}(x, s p^{-1-n}|p,q)p^{n-1}t^{n}. 
\end{eqnarray}
and the use of the Proposition \ref{ropo}. $\square$
\begin{remark} 
In the limit,
\begin{enumerate}
\item  When $(p,q)\to (1, 1)$ the  polynomials (\ref{pippo8}) are reduced to the
well known  Lucas polynomials
$L_{n}(x,s)$  given by the following formula \cite{natig}:
\begin{eqnarray}
\fl L_{n}(x,s|q):=\sum_{k=0}^{\lfloor n/2 \rfloor}
\frac{n}{n-k}{n-k\atopwithdelims()k}
s^{k}x^{n-2k}=
x^{\,n}\,{}_2F_1\left(\begin{array}{c}-\frac{n}{2}\,,\frac{1-n}{2}\\
1-n\end{array}\Bigg|-\frac{4s}{x^2}\right),\; n\geq0
\end{eqnarray}
 satisfying the following  recursion relation
\begin{eqnarray}
L_{n}(x,s|q) =  F_{n+1}
(x,s)+ s F_{n-1}(x, s)
\end{eqnarray}
and admitting the following generating function \cite{natig}
 \begin{eqnarray} 
f_L(x,s;t)=\sum_{n=0}^{\infty}F_{n}(x,s)t^{n}=\frac
{1+st^{2}}{1-xt-st^{2}},\quad |t|<1,
\end{eqnarray}
which can be easily derived from the  tree-term recursion relation (\ref{aas}) and  (\ref{geneq}).
For $s=1$, we recover the  normalized Lucas polynomials  
$l_{n}(x)=L_n(x,1)$ investigated by Bicknell \cite{Koshy}. 
\item 
For  $x=s=1$, the $(p,q)-$deformed Lucas polynomials   
become the
$(p,q)-$Lucas numbers: 
\begin{eqnarray}
L_{n}(p,q):= \sum_{k=0}^{\lfloor\,n/2\,\rfloor}
(p q)^{ ({}^k_2)}\frac{[n]_{p,q}}{[n-k]_{p,q}}\,{n-k\atopwithdelims[]k}
_{p,q}, 
\end{eqnarray}
 generalizing the $q-$Lucas numbers \cite{Cigl-I}
\begin{eqnarray}
L_{n}(q):= \sum_{k=0}^{\lfloor\,n/2\,\rfloor}
q^{ ({}^k_2)}\frac{[n]_{q}}{[n-k]_{q}}\,{n-k\atopwithdelims[]k}
_{q}. 
\end{eqnarray}
When $(p,q)\to (1,1)$ one obtains the well known Lucas number:
\begin{eqnarray}
\label{sama:numbe}
 L_n=\frac{1}{2^{n-1}}\sum_{k=0}^{n}{n
\atopwithdelims()2k }5^k.
\end{eqnarray} 
The $(p,q)-$generating function associated to the $(p,q)-$Lucas number is given by
\begin{eqnarray}
 f_L(t|p,q)=
\frac{ 1+ p t^2 }{1-p t}\,{}_2\varphi_2\left(\begin{array}{l}(p,q),\quad0\\
(p,t p q),(p,0)\end{array}
\Bigg|(p,q);-q t^2\right), \;|t|<1
\end{eqnarray}
  generalizing the $q-$Lucas number 
\begin{eqnarray}
 f_L(t|p,q)=
\frac{ 1+  t^2 }{1-t}\,{}_1\varphi_1\left(\begin{array}{c}q\\
t q\end{array}
\Bigg|q;-q t^2\right), \;|t|<1
\end{eqnarray}
while the classical  generating function $f_L( t)$ associated with the 
undeformed Lucas numbers  are given by
\begin{eqnarray}
f_L(t)=\sum_{n=0}^{\infty}L_{n}t^{n}=\frac
{1+t^{2}}{1-t-t^{2}},\quad |t|<1.         \end{eqnarray}
See \cite{StakhAran} for more details on  the Lucas numbers
$\{L_n\}_n$.
\end{enumerate}
\end{remark}
\begin{proposition}
\begin{eqnarray}
D_{(p,q)} L_n(n,s|p,q)=[n]_{p,q}F_n(x,s|p,q),\quad [n]_{p,q}=\sum_{k=0}^{n-1}p^{n-1-k}q^k.
\end{eqnarray}
\end{proposition}
 The proof stems from the observation that the
Fibonacci polynomials $F_n(x,s)$ (resp. the Lucas polynomials $L_n(x,s))$ are essentially the Chebyshev polynomials  of 
the second kind $U_n(x)$  (resp. of the first kind $T_n(x))$ (see  \cite{natig} for more details) with
$d_x T_n(x)=n U_{n-1}(x)$ \cite{ASK}.
\section{Fourier transforms of $F_n(x,s|p,q)$ and $L_n(x,s|p,q)$}
In this section, we derive 
explicit formulas for the classical Fourier integral transforms of  the
$(p,q)-$deformed Fibonacci $F_n(x,s|p,q)$ and  Lucas  $L_n(x,s|p,q)$ polynomials.
\subsection{Fourier transforms of the $(p,q)-$Fibonacci polynomials} 
Rewrite the $(p,q)-$deformed Fibonacci polynomials (\ref{pippo}) in the following form:
\begin{eqnarray}
 \label{pipp}
F_{n+1}(x,s|p,q)= \sum_{k=0}^{\lfloor\,n/2\,\rfloor}\,c_{n,\,k}
^{(F)}(p,q)\,s^{\,k}\,x^{\,n-\,2\,k}\,,                      \end{eqnarray}
where  the associated $(p,q)-$deformed coefficients  are given by
\begin{eqnarray}
 \label{coef}
c_{n,k}^{(F)}(p,q):= (p q)^{k(k+1)/2}{n-k\atopwithdelims[]k }_{p,q}.
\end{eqnarray}
Using the relations
\begin{eqnarray}
\label{sama:coef}
{n-k\atopwithdelims[]k }_{p^{-1},q^{-1}}=(p q)^{k(2k-n)}{n-k\atopwithdelims[]k }_{p,q}
\end{eqnarray}
 we can express  the $(p^{-1},q^{-1})-$deformed coefficients  from  (\ref{coef}) as 
\begin{eqnarray} 
\label{lab}
c_{n,k}^{(F)}(p^{-1},q^{-1})=(p q)^{-k(n+1-k)}c_{n,k}^{(F)}(p,q),
\end{eqnarray}
allowing to define the $(p^{-1}, q^{-1})-$Fibonacci polynomials in the form:
\begin{eqnarray}
 \label{inm}
\fl \qquad  F_{n+1}\big(x,s|p^{-1},q^{-1}\big):= \sum_{k=0}^{\lfloor n/2\rfloor}c_{n,k}
^{(F)}(p^{-1},q^{-1})s^{k}x^{n-2k}\cr
\fl \qquad  =x^{n}{}_{4}\varphi_3\left(\begin{array}{c}(p^{-\frac{n}{2}},q^{-\frac{n}{2}}),(p^{\frac{1-n}{2}},q^{\frac{1-n}{2}}),(p^{-\frac{n}{2}},-q^{-\frac{n}{2}}),(p^{\frac{1-n}{2}},-q^{\frac{1-n}{2}})
 \\\\
(p^{-n},q^{-n}),\quad 0,\quad0\end{array}\Bigg|
(p,q);-\frac{s}
{x^2}\right).
\end{eqnarray}
In the limit when $p\to1$, we immediately obtain the $q^{-1}-$Fibonacci polynomials as follows:

\begin{eqnarray}
\fl \qquad  F_{n+1}\big(x,s|q^{-1}\big):&=& \sum_{k=0}^{\lfloor n/2\rfloor}c_{n,k}
^{(F)}(q^{-1})s^{k}x^{n-2k}\cr
  &=&x^{n}{}_{4}\varphi_3\left(\begin{array}{c}q^{-\frac{n}{2}},q^{\frac{1-n}{2}},-q^{-\frac{n}{2}},-q^{\frac{1-n}{2}}
 \\\\
q^{-n},\quad 0,\quad0\end{array}\Bigg|
q;-\frac{s}
{x^2}\right).
\end{eqnarray}
The associated $q^{-1}-$Fibonacci number is given by 
\begin{eqnarray}
\fl \qquad  F_{n+1}\big(q^{-1}\big):&=& \sum_{k=0}^{\lfloor n/2\rfloor}c_{n,k}
^{(F)}(q^{-1})={}_{4}\varphi_3\left(\begin{array}{c}q^{-\frac{n}{2}},q^{\frac{1-n}{2}},-q^{-\frac{n}{2}},-q^{\frac{1-n}{2}}
 \\\\
q^{-n},\quad 0,\quad0\end{array}\Big|
q;-1\right).
\end{eqnarray}
\begin{theorem}
\label{propo1}
The Fourier   transform of the function $e^{-x^2/2}F_{n+1} (a e^{i \kappa x},s|p,q )$ is given by 
\begin{eqnarray}
\label{fourier}
\fl \qquad  \frac{1 }{\sqrt{2\pi}}\int_{\mathbb{R}}F_{n+1} (a e^{i \kappa x},s|p,q )e
^{ix y-x^2/2}d x=(p q)^{\frac{n^2}{4}}F_{n+1}
 \big(a e^{-\kappa y},p q s |p^{-1},q^{-1} \big) e^{-y^2/2}
\end{eqnarray}
leading to the formula
\begin{eqnarray}
F_{n+1}
 (a,s |p,q )=\frac{1}{ 2\pi }\int_{\mathbb{R}}\int_{\mathbb{R}}F_{n+1} \big(a e^{i\kappa x},s|p,q \big)e
^{ix y-x^2/2}d x d y,
  \end{eqnarray}
where $a$ is an arbitrary constant factor and $q=p^{-1}e^{-2\kappa^2}$.
\end{theorem}
 {\bf Proof.}
Using (\ref{pipp}) and (\ref{lab}), we obtain: 
\begin{eqnarray*}
& & \frac{1}{\sqrt{2\pi}}\int_{\mathbb{R}}F_{n+1}\big(a e^{i\kappa x},s|p,q\big)
e^{ix y-x^2/2}d x                     \\
& &=\sum_{k=0}^{\lfloor n/2\rfloor}c_{n,k}^{(F)}(p,q)s^{k}
a^{n-2k}\,\frac{1}{\sqrt{2\pi}}\int_{\mathbb{R}}e^{i x  y+ i(n- 2k)
 \kappa x- x^2/2} d x \\
& & =\sum_{k=0}^{\lfloor
n/2\,\rfloor}c_{n,k}^{(F)}(p,q)s^k a^{n-2k}e^{-\frac{1}{2}[\kappa (n-2k)+y]^2} \\       
& &= (p q)^{n^2/4}F_{n+1} \big(a e^{-\kappa y}
,p q s | p^{-1},q^{-1} \big)e^{-y^2/2},    
\end{eqnarray*}
where the Gauss integral transform $\int_{\mathbb{R}}e^{i x y-x^2/2} d x=
\sqrt{2\pi}e^{-y^2/2}$ is used.  The proof is achieved  by integrating   (\ref{fourier})   with respect to $y$. $\square$

In the limit when the parameter $p\to 1$, the Fourier   transform 
(\ref{fourier}) is reduced to the well-known results investigated by Atakishiyev et {\it al} \cite{natig},
\begin{eqnarray}
\frac{1 }{\sqrt{2\pi}}\int_{\mathbb{R}}F_{n+1} (a e^{i \kappa x},s|q )e
^{ix y-x^2/2}d x=q^{\frac{n^2}{4}}F_{n+1}
 \big(a e^{-\kappa y},q s |q^{-1} \big) e^{-y^2/2},
\end{eqnarray}
where $a$ is an arbitrary constant factor and $q= e^{-2\kappa^2}$.
\subsection{$(p,q)-$Lucas integral transform}
Rewrite here also the $(p,q)-$deformed Lucas polynomials  (\ref{pippo8})   as
\begin{eqnarray}
L_{n}(x,s|p,q)= \sum_{k=0}^{\lfloor n/2\rfloor}
c_{n,k}^{(L)}(p,q)s^{k}x^{n-2k},
\end{eqnarray}
where  the coefficients  $c_{n,k}^{(L)}(p,q)$ are given by
\begin{eqnarray}
\label{coefs}
c_{n,k}^{(L)}(p,q):= (p q)^{k(k-1)/2}\frac{[n]_{p,q}}{[n-k]_{p,q}}
{n-k\atopwithdelims[]k }_{p,q}.               
\end{eqnarray}
By using (\ref{sama:coef}),
 one can show   that the $(p^{-1},q^{-1})-$coefficients   (\ref{coefs}) can be expressed as:
\begin{eqnarray} 
c_{n,k}^{(L)}(p^{-1},q^{-1})=(p q)^{k(k-n)}c_{n,k}^{(L)}(p,q)
\end{eqnarray}
permiting to define   the $(p^{-1}, q^{-1})-$deformed Lucas polynomials  as follows:
\begin{eqnarray}
 \label{inm}
\fl \qquad  L_n\big(x,s|p^{-1},q^{-1}\big):= \sum_{k=0}^{\lfloor n/2\rfloor}c_{n,k}
^{(L)}(p^{-1},q^{-1})s^{k}x^{n-2k}\cr
\fl \; =x^{n}{}_{4}\varphi_3\left(\begin{array}{c}(p^{-\frac{n}{2}},
q^{-\frac{n}{2}}),(p^{\frac{1-n}{2}},q^{\frac{1-n}{2}}),
(p^{-\frac{n}{2}},-q^{-\frac{n}{2}}),(p^{\frac{1-n}{2}},-q^{\frac{1-n}{2}})
 \\\\
(p^{1-n},q^{1-n}),\quad 0,\quad0\end{array}\Bigg|
(p,q);-\frac{s p q}
{x^2}\right).
\end{eqnarray}
The limit when $p\to1$
yields the $q^{-1}-$Lucas polynomials \cite{natig}:
\begin{eqnarray}
\fl  L_n\big(x,s|q^{-1}\big)= \sum_{k=0}^{\lfloor n/2\rfloor}c_{n,k}
^{(L)}(q^{-1})s^{k}x^{n-2k} 
 =x^{n}{}_{4}\varphi_3\left(\begin{array}{c}q^{-\frac{n}{2}},q^{\frac{1-n}{2}},-q^{-\frac{n}{2}},-q^{\frac{1-n}{2}}
 \\\\
q^{1-n},\quad 0,\quad0\end{array}\Bigg|
q;-\frac{s  q}
{x^2}\right).
\end{eqnarray}
Their associated $q^{-1}-$Lucas numbers are found when $x=s=1$ as:
\begin{eqnarray}
  L_n\big(q^{-1}\big)
 = {}_{4}\varphi_3\left(\begin{array}{c}q^{-\frac{n}{2}},q^{\frac{1-n}{2}},-q^{-\frac{n}{2}},-q^{\frac{1-n}{2}}
 \\\\
q^{1-n},\quad 0,\quad0\end{array}\Bigg|
q;- q\right).
\end{eqnarray}
Finally, we have the following:
\begin{theorem}
The Fourier  transform of the function $e
^{-x^2/2}L_{n} (be^{i \kappa x},s|p,q )$ is given by
\begin{eqnarray}
\label{sama:fouriertrans}
\fl  \qquad 
\frac{1}{\sqrt{2\pi}}\int_{\mathbb{R}}L_{n} (be^{i \kappa x},s|p,q )e
^{i x y-x^2/2}d x=(p q)^{\frac{n^2}{4}}L_{n}
 \big(be^{-\kappa\,y},(p q)^{-1}s\, |p^{-1},q^{-1} \big)e^{- y^2/2}  
\end{eqnarray}
providing the formula
\begin{eqnarray}
L_{n}
 (b,s |p,q )=\frac{1}{ 2\pi }\int_{\mathbb{R}}\int_{\mathbb{R}}L_{n} (b e^{i\kappa x},s|p,q )e
^{i x y-x^2/2}d x d y,  
\end{eqnarray}
where $b$ is an arbitrary constant 
factor and $q=p^{-1}e^{-2\kappa^2}$.
\end{theorem}
{\bf Proof.}  The proof is immediate 
from the Theorem \ref{propo1}. $\square$

In the limit when the parameter $p\to 1$, the
 Fourier   transform (\ref{sama:fouriertrans}) is reduced 
to the well-known formula derived by
 Atakishiyev et {\it al} \cite{natig}
\begin{eqnarray}
\frac{1}{\sqrt{2\pi}}\int_{\mathbb{R}}L_{n} (be^{i \kappa x},s|q )e
^{i x y-x^2/2}d x=q^{\frac{n^2}{4}}L_{n}
 \big(be^{-\kappa\,y},q^{-1}s\, | q^{-1} \big)e^{- y^2/2},
\end{eqnarray}
where $a$ is an arbitrary constant
 factor and $q= e^{-2\kappa^2}$.
\section{Conclusion}
In the present work, a full characterization of 
$(p,q)-$deformed Fibonacci and Lucas
polynomials  has been achieved. 
These polynomials obey  non-conventional three-term
 recursion relations as previously shown for their $q-$analogs. 
Besides,  the formulae for the computation of the  associated  Fourier 
integral transforms have been deduced. Previous known results have been 
recovered as particular cases and properly discussed. 

\section*{Acknowledgements}
This work is partially supported by the Abdus Salam International
Centre for Theoretical Physics (ICTP, Trieste, Italy) through the
Office of External Activities (OEA) -\mbox{Prj-15}. The ICMPA
is in partnership with
the Daniel Iagolnitzer Foundation (DIF), France.

\end{document}